\documentclass[preprint,12pt]{elsarticle}

\usepackage{graphicx}

\usepackage{amssymb}

\newcommand{\snn}{$\sqrt{s_{_{\rm NN}}}$}

\journal{Nuclear Physics A}

\begin{document}

\begin{frontmatter}


\title{Dependence of Single Particle Distributions on Rapidity and Centrality in d+Au Collisions at \snn = 200 GeV }
\author{R. Debbe BNL }

\begin{abstract}
 Measurements of identified single particle distributions in d+Au collisions at \snn = 200 GeV at the RHIC
collider are described. The dependence of these distributions on the centrality of the collisions, as well 
as the rapidity of the detected particles are emphasized in this report.
   
\end{abstract}


\end{frontmatter}


\section{Introduction}

{\bf Early p+A experiments.} Collisions of proton beams incident on fixed nuclear targets  have been used in the past 
to study
hadronic interactions. The energies of the incident beams ranged from a few GeV  
to over 800 GeV. Energy loss and baryon transport in nuclear matter were 
expected to provide a clear window into the workings of strong interactions at the time
when Quantum Chromodynamics (QCD) was being formulated. The puzzling image that emerged from these early 
studies did not show evidence for intranuclear cascade but rather suggested that the 
projectile interacted at several points within the nuclear target
receiving a transverse momentum ``kick'' at each interaction before undergoing a final
violent interaction with a target nucleon. 
Particle yields from p+A collisions increase with the size of the target and follow a power law dependence
in atomic number A: $d\sigma_{pA->hX}/d^{2}p_{T} = d\sigma_{pp->hX}/d^{2}p_{T}A^{\alpha(p_{T})}$ \cite{Cronin}. 
The majority of the 
detected particles were found on the target side of the collision \cite{Busza:1975fe, Busza:1989px}.

{\bf The Cronin enhancement.} Those early experiments on nuclear targets compared particle production as function of the transverse 
momentum, to the base-line 
p+p collisions at the same energy
in order to highlight differences related to nuclear effects.
In general, the comparison would be done between two A and B target systems and it assumes scaling
with the number of binary collisions $N_{coll}$ (incoherent sum of nucleon-nucleon interactions). 
This comparison
is now known as the nuclear modification factor and for d+Au collisions compared to p+p interactions is written as:
 $R_{dAu} = (dN_{dAu}/dydp^{2}_{T})/ (N^{dAu}_{coll}dN_{pp}/dydp^{2}_{T})$. 
At values of $p_{T}$ close to zero, the $R_{dAu}$ 
is smaller than 1 and appears to scale with the number of participant nucleons $N_{part}$. This may be an
indication that this behavior is driven by the still unresolved non-perturbative QCD environment. 
At intermediate values of $p_{T}$ (from 1 to 4 GeV/c), an enhancement
has been measured, which  is commonly related to multiple interactions of the projectile in the target 
that do not produce particles but excite the projectile wave function. This excitation is postulated 
to be a hardening  of the transverse momentum distribution of the projectile partons. At a later time,
the excited projectile will fragment and the detected particle will ``remember'' the transverse parton
distribution of the beam.  

{\bf The DIS program.} A new aspect of the p+A collisions resulting from theoretical advances in QCD and experimental results 
 probing deep into the structure of nucleons has gained much relevance in the last 10 to 20 years
and is usually referred to as small-x physics.  
The HERA Deep Inelastic Scattering (DIS) program has explored the structure of protons with unprecedented 
reach. The QCD based image that comes out of that program shows nucleons in x (the longitudinal momentum 
fraction carried by a parton) and $Q^{2}$ (the magnitude of the 4-vector momentum of the probe, a virtual
photon in the case of DIS which determines the probing resolution).
At high values of x the parton distribution is dominated by the constituent u, d and s quarks.
As the values of x tends toward zero, the population of partons becomes a mixture of gluons and quark-antiquark 
pairs that emerge as fluctuations from the vacuum \cite{:2009wt}. At the limit of even smaller values of x, gluons 
are the dominant entities, and the HERA measurements found that the gluon density grows as a power in $1/x$ \cite{Breitweg}.
An end to that growth has not yet been measured, but unitarity imposed on the wave function of the target 
makes the end of that growth all but inevitable.
The growth of the gluon density is postulated to have two components: a linear term produced by gluon splitting,
and a non-linear one related to gluon fusion as expressed in eq. \ref{eq:QuantumEvol}:

\begin{equation}
dG/dy \sim G - G^{2}.
\label{eq:QuantumEvol}
\end{equation} 

As the value of the momentum fraction x in the target reaches sufficiently small values and the growth
of the gluon density  
governed by eq. \ref{eq:QuantumEvol} has ended, the target is said to be in the so called   Color Glass 
Condensate (CGC) a slowly varying system of color fields that presents a maximal cross section to any
projectile \cite{Gelis:2010nm} . A boundary to the CGC is expected to depend on $1/x$ as well as the size 
of the target A:

\begin{equation}
Q^{2}_{sA} = A^{1/3}Q^{2}_{0}(x_{0}/x)^{\lambda} = A^{1/3}Q^{2}_{0}x_{0}^{\lambda}e^{\lambda y}.
\label{eq:SatScale}
\end{equation} 

When one describes p+A collisions at the partonic level, the momentum fraction of the target parton depends on the
rapidity of the produced particle, and for simplicity, it is written as $x_{2}= (2p_{T}/\sqrt{s})e^{-y}$. In the 
CGC framework it is possible to search for the onset of saturation ($p_{T}$ below or close to $Q^{2}_{sA}$) 
by studying
particle production as function of rapidity and centrality \cite{Baier:2003hr, Kharzeev:2004yx}. 
The High Rapidity section of this report is an attempt to summarize the measurements performed in d+Au 
collisions with relevance to small-x physics.
 
{\bf The RHIC d+Au program.} 
The d+Au collisions at RHIC were in part included in the program as a contrasting system to the hot 
medium being
formed in Au+Au collisions at \snn = 200 GeV. The first result fron d+Au at RHIC did not show the 
suppression
found in Au+Au in what is understood as the absence of ``jet quenching'' in cold matter. Instead the nuclear
modification factor measured at mid-rapidity brought back the focus on the Cronin effect, which later was 
found to be different when measured for baryons and mesons. The yields of charged particles were found
to be suppressed in the projectile fragmentation region in a manner consistent with the onset of 
saturation in the Au wave function.

\begin{figure}[ht]
\begin{minipage}[b]{0.47\linewidth}
\centering
\includegraphics[width=2.7in]{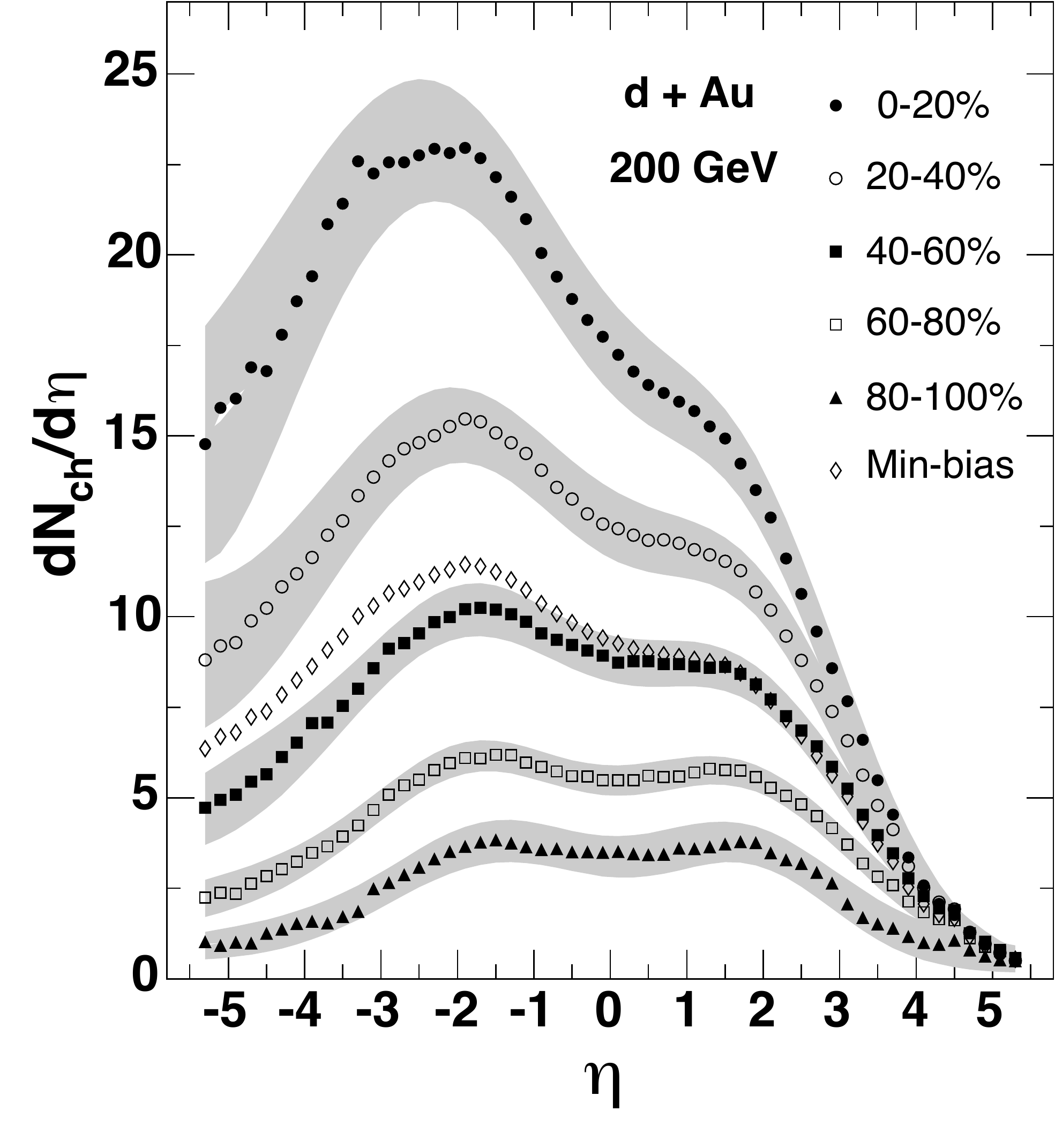}
\caption{Pseudo-rapidity distributions of charged particles measured in d+Au collisions at RHIC  
\snn=200 GeV for data samples of different centrality as well as a Minimum Bias sample \cite{PhobosDndeta}.}
\label{Fig:PhobosDndeta}
\end{minipage}
\hspace{0.5cm}
\begin{minipage}[b]{0.47\linewidth}
\centering
\includegraphics[width=2.7in]{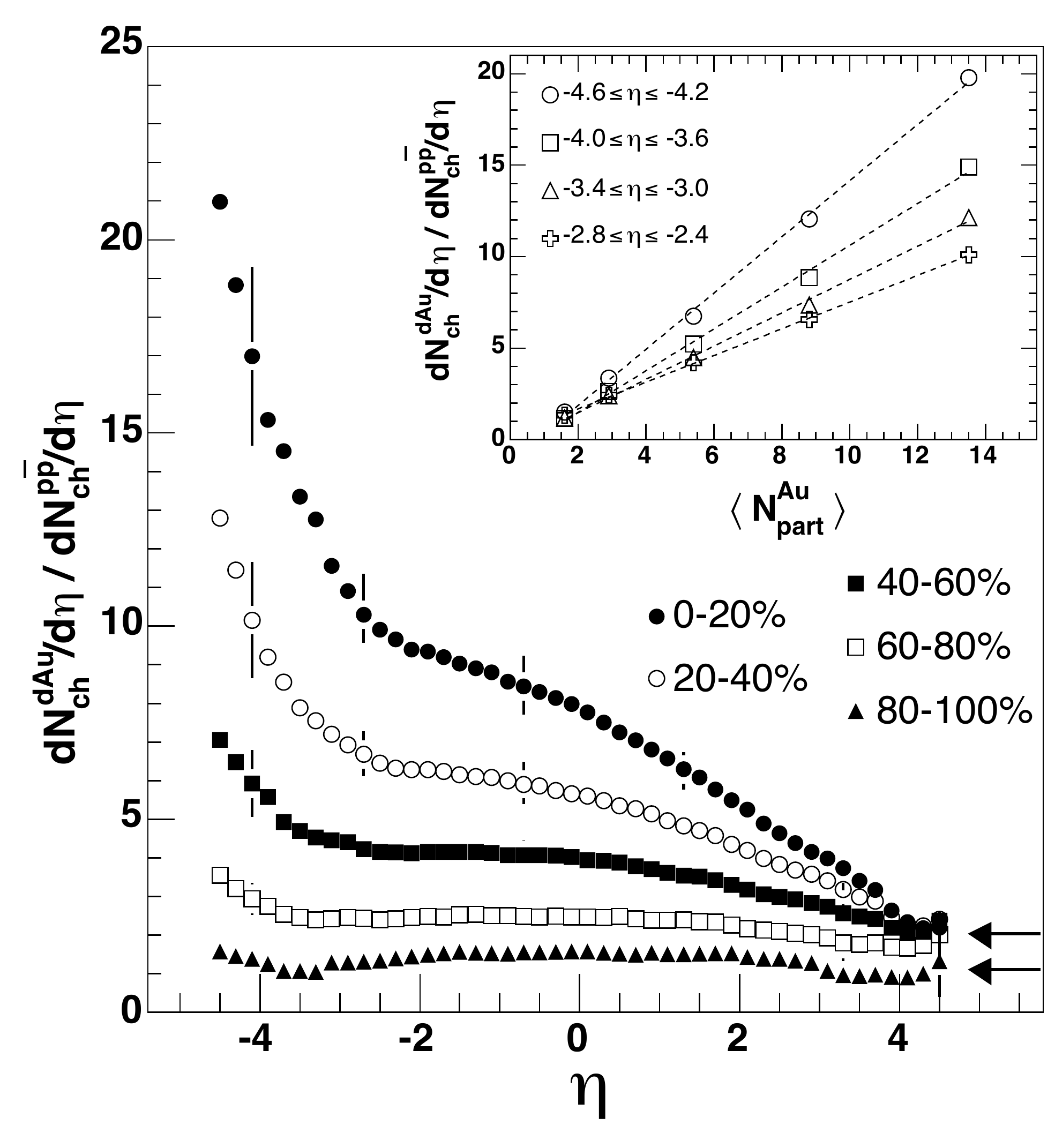}
\caption{Centrality dependence of the $dN_{ch}/d\eta$ ratio of d+Au collisions relative to that extracted 
from inelastic p + p collisions at the same energy by the PHOBOS Collaboration.}
\label{Fig:triangular}
\end{minipage}
\end{figure}

At RHIC, d+Au and p+p collisions at \snn = 200 GeV were, for the  first time in p+A physics history,
collected in collider mode. At top RHIC energy these collisions provide well separated fragmentation 
regions, and as such, constitute a leap in data quality and physics reach. Figure \ref{Fig:PhobosDndeta}
shows the pseudo-rapidity distribution of charged particles measured in d+Au collisions at RHIC
by the PHOBOS Collaboration \cite{PhobosDndeta}.

The pseudo-rapidity density of the very early p+A experiments suggested a projectile traversing the
target without being affected. This feature of the data was dubbed  ``nuclear transparency'' in p+A collisions.
The RHIC results are showing a much clearer picture that negates the existence of such transparency.
When compared to similar pseudo-rapidity distribution extracted from p+p collisions at the same energy, as is
done if Fig. \ref{Fig:triangular}, one finds a triangular shape extending from $\eta=0$ to close to the deuteron 
beam rapidity, which can be explained by a scenario where each detected particle ``drags'' uniform trails of
particles at lower rapidities as suggested by Brodsky, Gunion, and K\"{u}hn \cite{Brodsky:1977de}. 
The particle production on the target
fragmentation side ($\eta<0$) is completely different; as it was the case at lower energies, most  of the 
particle yield detected originates from that region and when compared to p+p yields, one finds a clear change 
in the growth rate, deeper into the target fragmentation ($\eta<-3$).

The RHIC d+Au collisions have been studied with a complement of four experiments: STAR with its
 full azimuth coverage and its large acceptance Time Projection Chamber (TPC) which covers 2 units 
in pseudo-rapidity ($|\eta|<1$).
PHENIX, the second large detector at RHIC,
designed with emphasis on electron, photon and muon detection. PHOBOS, with its almost complete 
coverage for measurements of charged particle multiplicity, and finally, BRAHMS with its two conventional rotatable
spectrometers that measure fully identified charged particles starting from mid-rapidity and 
extending well into the fragmentation region of the deuteron beam.
These four experiments collected their data under similar trigger conditions.
STAR defined its minimum-bias trigger by requiring at least one
beam neutron in the Au side Zero Degree Calorimeter (ZDC). (The ZDC counters are located at 18 meters on both 
sides of the nominal interaction point).  The Beam-Beam Counter (BBC) ($3.3<|\eta|<5.0$) was used to measure 
the luminosity and to define the vertex of the collisions. 
The STAR minimum-bias trigger measures $95\pm3\%$ of hadronic d+Au cross section ($2.21\pm0.09$b).
The PHENIX minimum-bias trigger was defined by hits on their BBC detector ($3.0<|\eta|<3.9$) and a vertex cut: 
 $|z|<30$ cm. This trigger  accepts $88\pm4\%$ of all
inelastic d+Au that satisfy the vertex condition. 
PHOBOS triggers with 16 scintillator paddles installed with azimuthal symmetry and cover $3<|\eta|<4.5$.
The BRAHMS minimum-bias trigger was defined with plastic scintillators placed 
at $\pm1.6$, $\pm4.2$ and $\pm 6.6$ meters from the nominal interaction point, which cover
$3.2 < |\eta| <5.3$ and can trigger on $91\pm3\%$ of the d+Au inelastic cross section 
(2.4b).
The centrality of the STAR events  is defined with raw (uncorrected) multiplicity of charged particles 
measured in the Au side Forward TPC (FTPC-Au) ($-3.8<\eta<-2.8$).
PHOBOS used the signal from their three ring detectors ($3.0<|\eta|<5.4$).
PHENIX defines centrality with the charged particle multiplicity in the Au side BBC. 
BRAHMS defines centrality with  Si strips and scintillator tile multiplicity counters ($-2<\eta<2$).

\section{Mid-rapidity}

\begin{figure}[ht]
\begin{minipage}[b]{0.5\linewidth}
\centering
\includegraphics[width=3.0in]{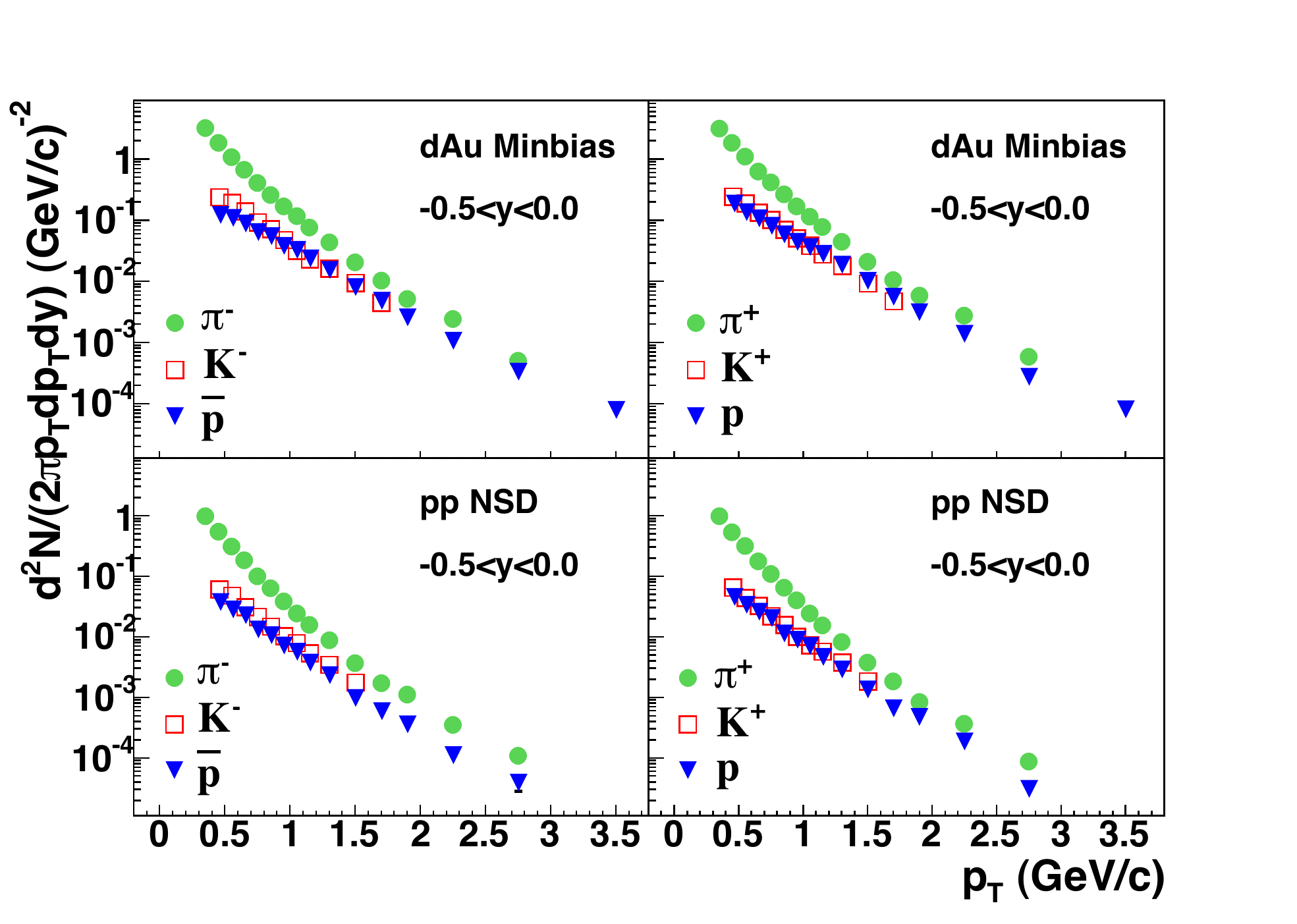}
\caption{STAR identified charged particle yields extracted from a Minimum bias sample of d+Au 
collisions at \snn = 200 GeV as well as Non Single Diffractive (NSD) p+p at the same energy \cite{:2008ez}. }
\label{Fig:StarSpectra}
\end{minipage}
\hspace{0.5cm}
\begin{minipage}[b]{0.47\linewidth}
\centering
\includegraphics[width=2.7in]{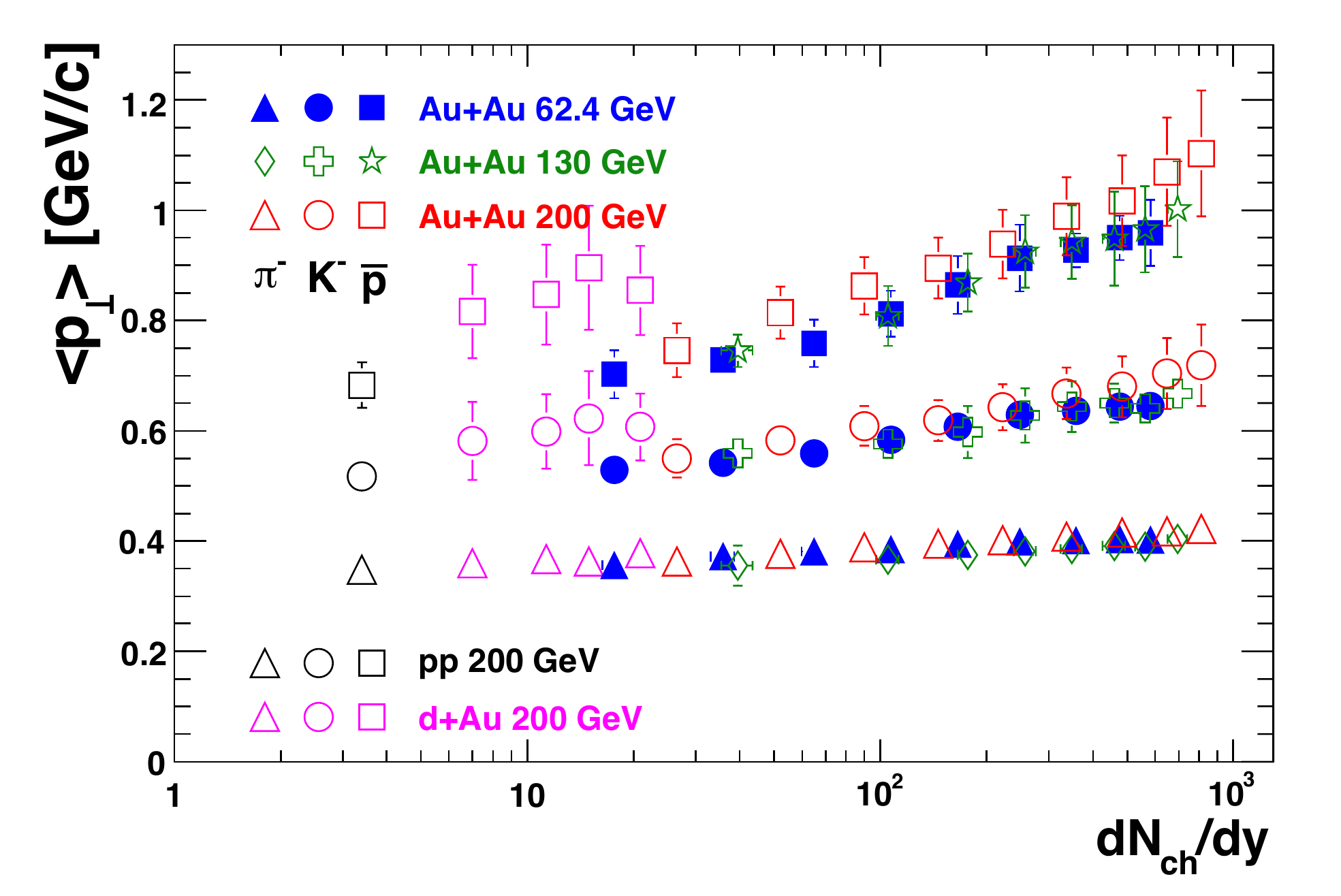}
\caption{(color online) Average $p_{T}$ for $\pi$ K and protons as function of $dN_{ch}/dy$ in p+p d+Au and 
Au+Au collisions at different centralities. Statistical and systematic error added in quadrature.  }
\label{Fig:meanPt}
\end{minipage}
\end{figure}

The STAR Collaboration has produced a 
systematic study of particle production at mid-rapidity with moderate transverse momentum reach 
from all collision systems  
studied at RHIC so far; p+p, d+Au and Au+Au. In particular, the invariant yields for identified charged particles
produced in minimum-bias d+Au and p+p collisions at the same energy are shown in Fig. \ref{Fig:StarSpectra} 
as a sample of the high quality of the data. In order to compare the behavior of these system, they 
performed fits to particle spectra based in the Blast Wave formalism, even if the smaller systems p+p 
and d+Au are not expected to show radial flow. The average transverse momentum $\langle p_{T} \rangle$ of pions 
displayed as a function of charged particle density in pseudo-rapidity space in Fig. \ref{Fig:meanPt} shows 
a smooth linear growth spanning
from p+p all the way to the most central Au+Au collisions. In contrast, the corresponding quantity 
extracted for kaons and protons shows values for p+p and d+Au that are visibly higher than the corresponding 
extrapolations from Au+Au collisions as they approach the most peripheral collisions. The most peripheral 
Au+Au collisions have the same value of $\langle p_{T} \rangle$ as the p+p collisions. As stated in the STAR 
publication, 
the different behavior for kaons and protons in d+Au collisions may come from jet contributions (in Au+Au jet 
contribution is softened by quenching) or initial state multiple interactions in the Au target.

\begin{figure}[ht]
\begin{minipage}[b]{0.5\linewidth}
\centering
\includegraphics[width=3.0in]{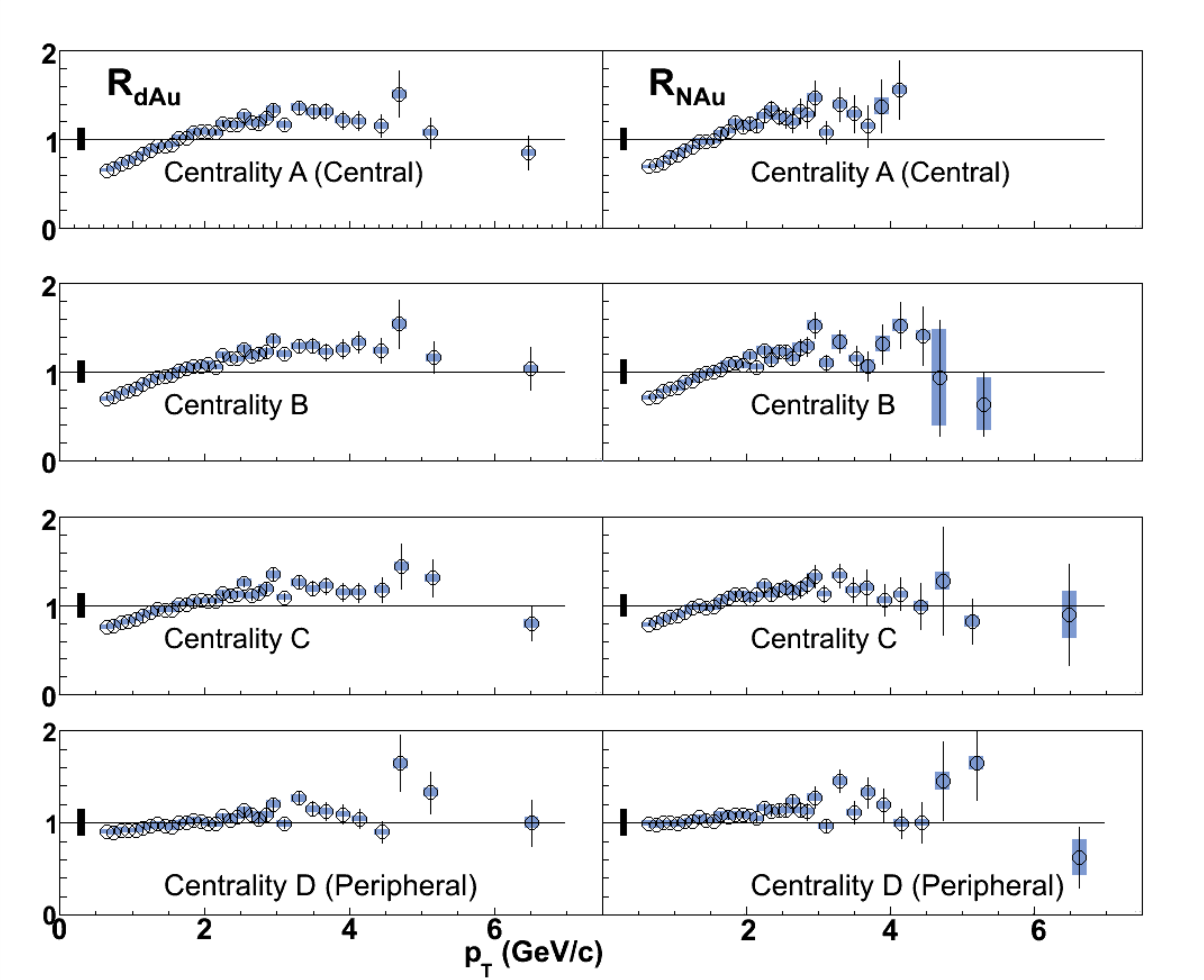}
\caption{PHENIX $R_{dAu}$ and $R_{NAu}$ for charged hadrons in four different centrality 
samples A: 0-20\%, B: 20-40\%, C: 40-60\%, D: 60-88\% of d+Au cross-section \cite{:2007by}.}
\label{Fig:RdAuPhenixHadron}
\end{minipage}
\hspace{0.5cm}
\begin{minipage}[b]{0.47\linewidth}
\centering
\includegraphics[width=2.7in]{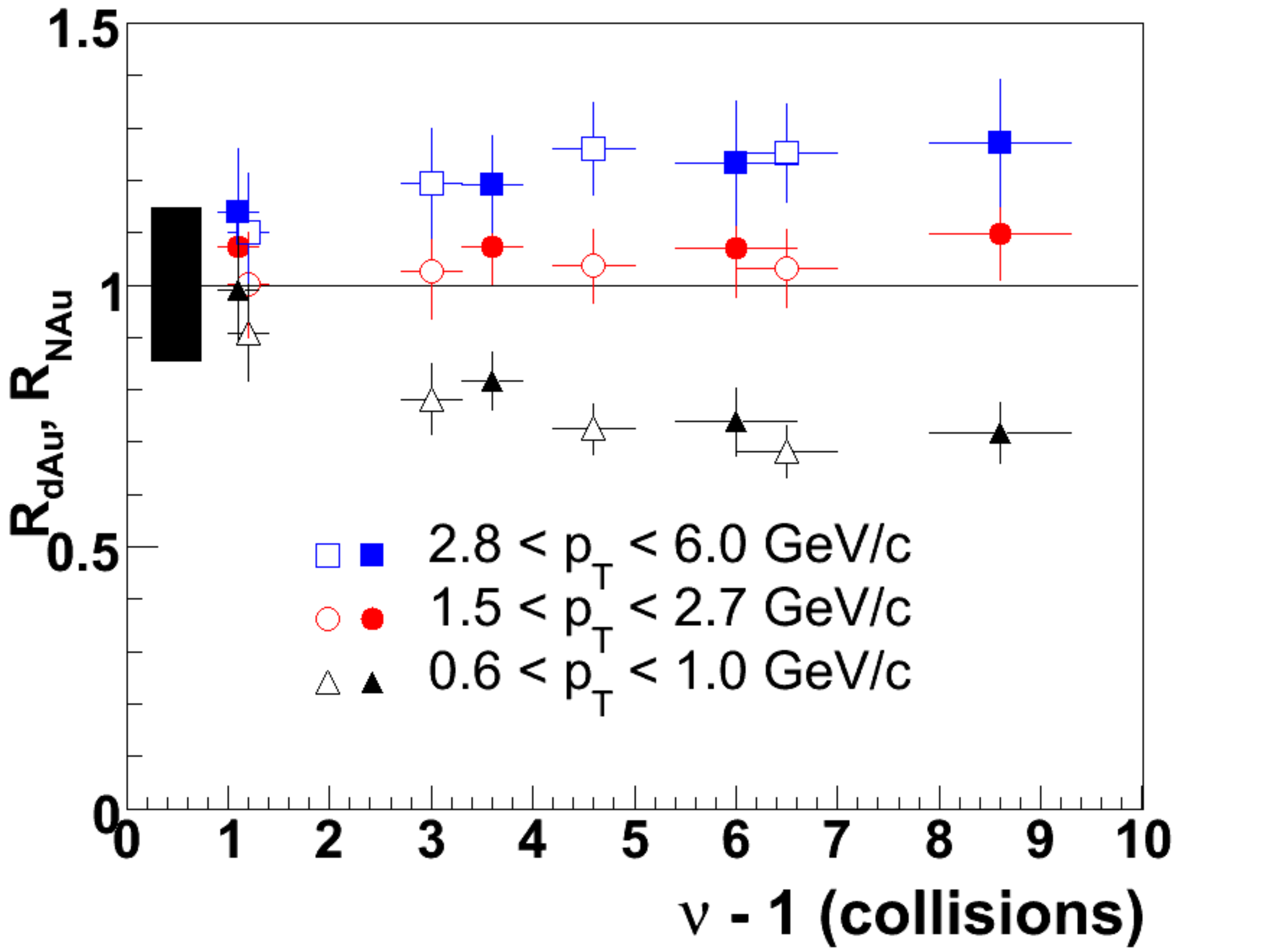}
\caption{RdAu (open symbols) and $R_{NAu}$ (closed symbols) for charged particles in three momentum 
ranges as function of $\nu - 1$ \cite{:2007by}.  }
\label{Fig:Nudependence}
\end{minipage}
\end{figure}

The PHENIX Collaboration has performed a study of the Cronin effect at mid-rapidity with charged particles,
both with the full d+Au collision events as well as events where a beam neutron was detected in the deuteron side  ZDC 
or a beam 
proton in the Forward Calorimeter North, which is also located in the deuteron fragmentation
region. These tagged  events divide into two samples where only one nucleon in the deuteron beam
interacted: p+Au or n+Au. In what follows both samples are used together and those collisions are 
designated as nucleon+Au (N+Au). 
 The nuclear modification factors $R_{dAu}$ and $R_{NAu}$ for different centralities are shown in Fig. 
\ref{Fig:RdAuPhenixHadron}. The centrality of the event is defined with the multiplicity of charged 
particles incident in the BBC (in the Au fragmentation region) which is a related to the number of 
participant 
Au target nucleons. The N+Au sample of collisions is biased towards peripheral collisions but
the number of Au participant nucleons can be extracted from a Glauber model Monte Carlo calculation used
as well to extract collision parameters for the d+Au collisions. The similarity of the $p_{T}$ dependence 
of the $R_{dAu}$ and $R_{NAu}$ factors is clear and it may be related to the fact that the Cronin effect
is tied to the partonic component of the nucleons.

Figure \ref{Fig:Nudependence} is a summary of the previous figure where an average of the nuclear 
modification factors $R_{dAu}$ and $R_{NAu}$ in three ranges of transverse momentum is shown as 
function of the number of sequential nucleon-nucleon collisions. 
A QCD based description of the Cronin effect has a projectile nucleon interacting several times with
the nuclear target, each interaction broadens the transverse momentum distribution of the projectile 
partons. Eventually, the projectile has a hard interaction or hadronizes into other particles whose 
transverse momentum distribution will display the history of the previous interactions. Most of the 
models describing Cronin effect will parametrize the target contribution to the $k_{T}$ broadening 
as proportional to the number of sequential interactions minus one $\nu -1$. 
Where $\nu = <N_{coll}/N^{d}_{part}>$ in d+Au collisions and $\nu = N_{coll}$ for N+Au.  
The Cronin effect is most pronounced in the 
high $p_{T}$ range, where, within the systematic uncertainties, it reaches a constant value of 
$\sim 20\%$ above $\nu=5$ and remains flat for more central collisions. Both d+Au and N+Au have the 
same $\nu$ 
dependence.

\begin{figure}[ht]
\begin{minipage}[b]{0.47\linewidth}
\centering
   \includegraphics[width=2.9in]{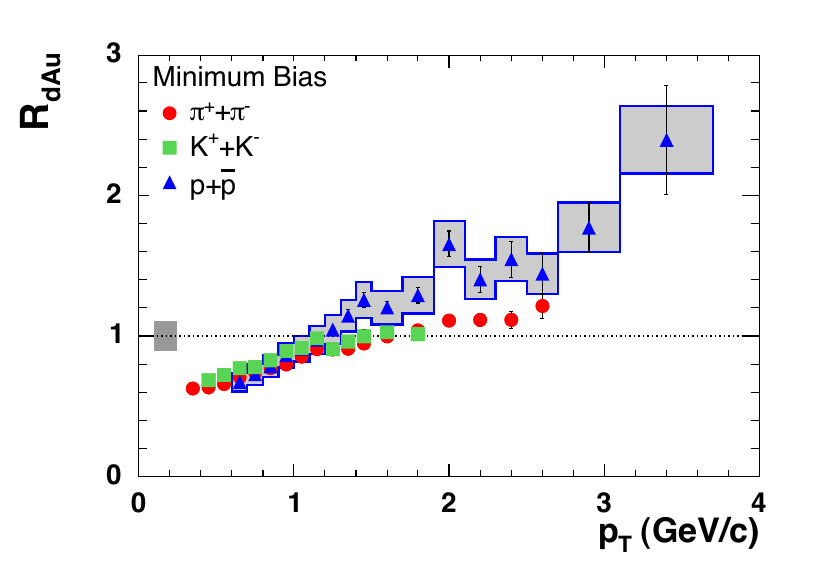}
  \caption{(color online)  The nuclear modification factor $R_{dAu}$ extracted from
identified charged particles at mid-rapidity from a minimum bias sample of d+Au collisions 
\cite{PhenixNucEff} (PHENIX). }
\label{Fig:phenixNuclearEff}
\end{minipage}
\hspace{0.5cm}
\begin{minipage}[b]{0.47\linewidth}
\centering
   \includegraphics[width=2.2in]{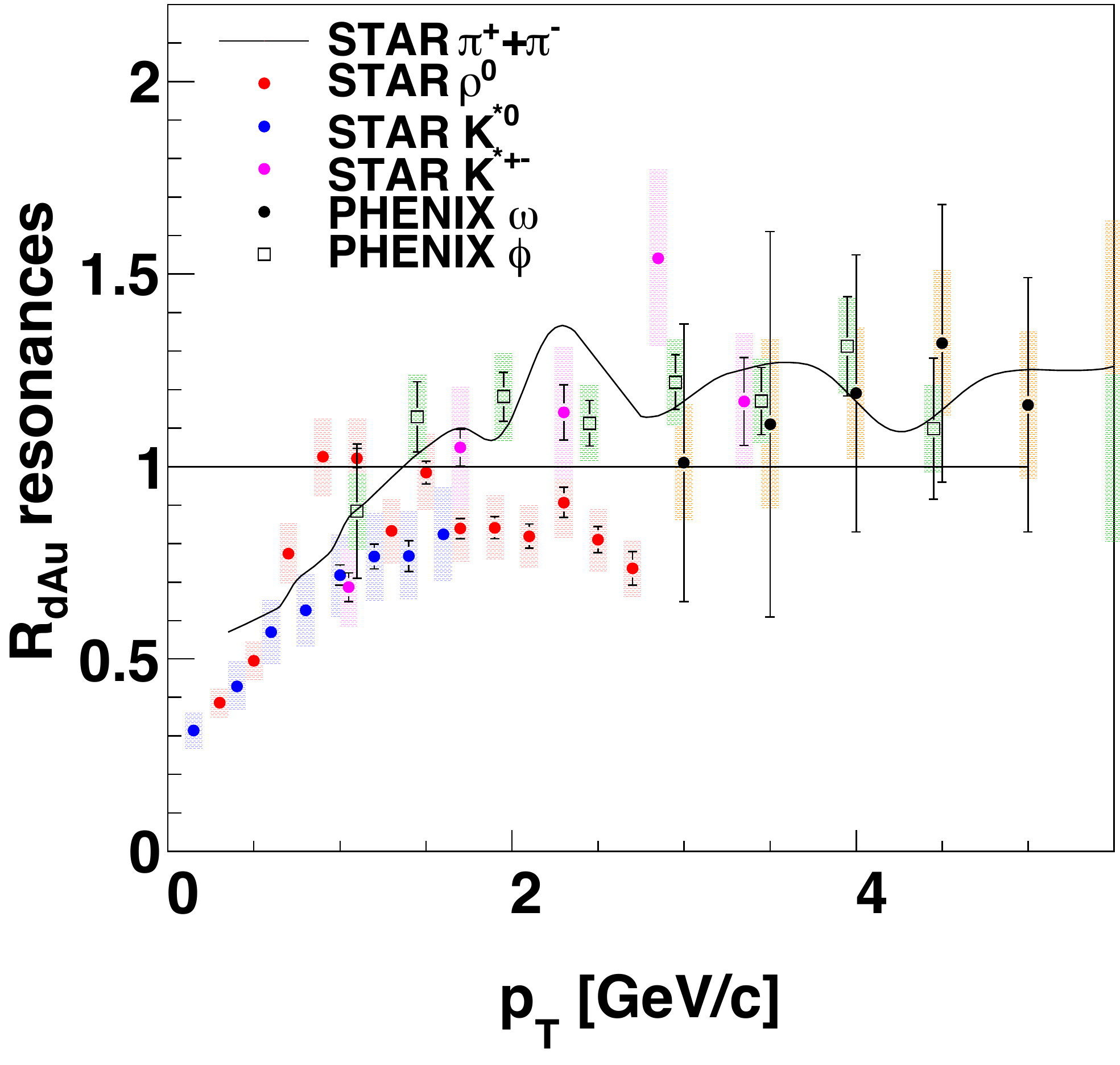}
  \caption{(color online) The $R_{dAu}$ for resonances measured by STAR and PHENIX. For comparison, 
the same ratio
is displayed for charged pions. }
\label{Fig:resonances}
\end{minipage}
\end{figure}

The nuclear modification factors extracted with identified particles have shown a different behavior 
between baryons and mesons. Figure \ref{Fig:phenixNuclearEff} shows that the $R_{dAu}$ for protons deviates 
from binary collisions scaling by as much as 40\% whereas pions and kaons show an enhancement of at 
most 10\%. 
The difference in the magnitude of the Cronin effect between protons and pions has been seen at lower 
energies \cite{Antresian}, and it is stronger that the one measured at RHIC energies . 
This difference  continues to be unexplained and both the STAR and PHENIX collaborations have published several
studies that add information about the Cronin effect and its difference between baryon and mesons.  Figure  
\ref{Fig:resonances} compiles those studies 
showing the $R_{dAu}$ ratio constructed for $\rho$, $K^{*\pm}$ and $K^{*0}$ extracted from d+Au and p+p collisions 
at mid-rapidity by STAR 
\cite{Abelev:2008yz}, and the   $\phi$ \cite{PhenixPhi} and $\omega$ \cite{PhenixOmega} mesons studied by PHENIX. 
The figure includes a curve joining the measured STAR $R_{dAu}$ for charged pions. Within errors, it appears as all
measured mesons behave just as pions, even though the $\rho$ tends to have smaller values above 2 GeV/c.  

\section{High rapidity}

\begin{figure}[!ht]
   \includegraphics[height=4.in]{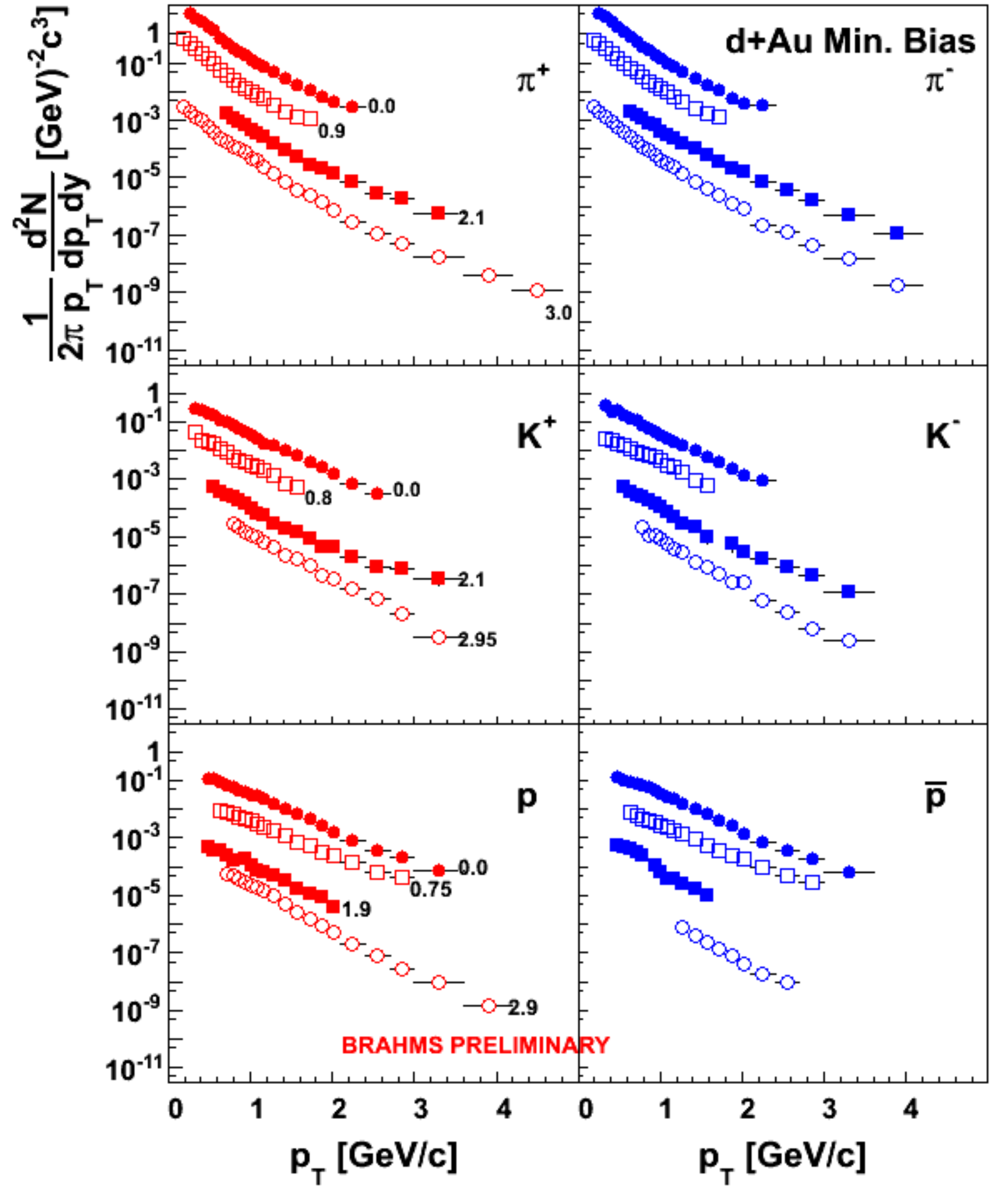}
  \caption{(color online) Invariant yields for positive pions, kaons and protons are shown in 
the left panels with
open and filled symbols at four rapidity values indicated at the high end of each distribution. The 
corresponding negative particle spectra are shown on the right panels with open and filled symbols.
For clarity, the distributions at different rapidities are shifted by a factor of 10.  }
\label{Fig:brahmsSpectra}
\end{figure}

The BRAHMS experimental setup consists of two conventional rotatable spectrometers. The Mid-Rapidity 
Spectrometer (MRS) tracks charged particles from 90 to 35 degrees with two Time-Projection Chambers and 
two Time-of-Flight detectors. The Forward Spectrometer (FS) measures charged particles produced at angles
ranging from 30 to 2.3 degrees. As the momenta of those particles can reach values as high as 30 GeV/c,
the momentum measurement is done with a complement of four magnets, two TPCs and two 
drift chambers. Particle identification at high momentum is done with the Ring Imaging Cherenkov detector.
Lower momentum charged particles are identified using two Time-of-Flight detectors.
Figure \ref{Fig:brahmsSpectra} shows the minimum-bias yields for positive and negative pions, kaons and 
protons measured 
with the MRS and FS spectrometers in d+Au collisions at \snn = 200 GeV. These distributions are fully
corrected for spectrometer acceptance, decays in flight and absorption in the spectrometers' material.
The largest systematic uncertainty
in these results is related to the matching between magnetic field settings at any angular setting of
the spectrometers and is estimated to be have an upper value of 10\%.  
 
\begin{figure}[htb]
   \includegraphics[height=2.5in]{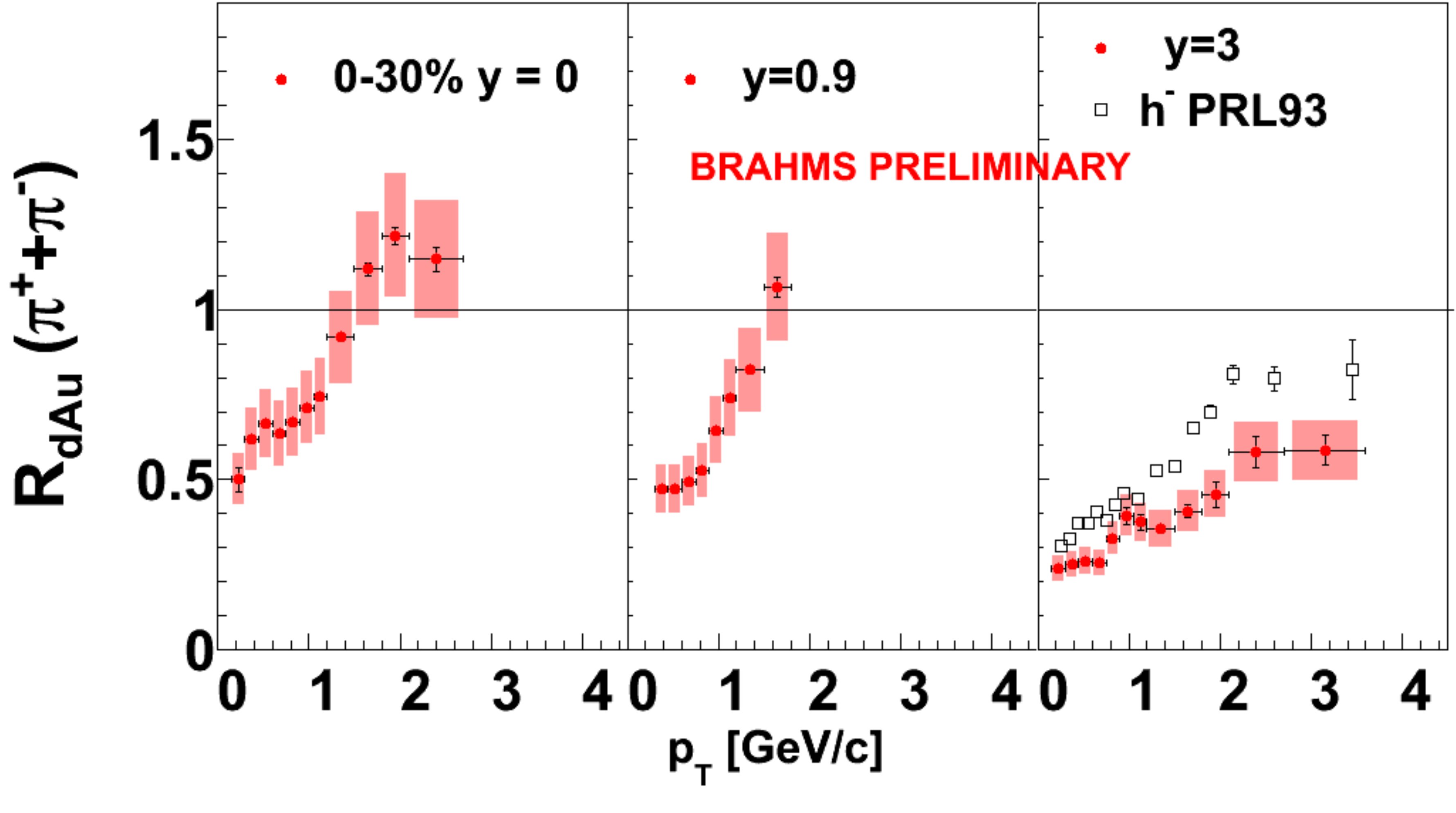}
  \caption{ The nuclear modification factor $R_{dAu}$ for charged pions detected in the most central
sample of events (0-30\%) at three different
values of rapidity. The reference p+p charged pion distributions were also measured with the BRAHMS
spectrometers during RHIC run 5. At y=3, the published $R_{dAu}$ for negative charged hadrons 
\cite{Arsene:2004ux}, is shown with open
square symbols for comparison. }
\label{Fig:brahmsRdAuPion}
\end{figure}

 Figure \ref{Fig:brahmsRdAuPion} shows the nuclear modification factor 
$R_{dAu}$ comparing the invariant yields of  charged pions in 
the most central sample of d+Au collisions at \snn = 200 GeV ($N_{coll}$ = 12.6),  to similar 
yields measured in p+p collisions at the same energy collected with the BRAHMS spectrometers. The 
two panels to the left of the figure show the $R_{dAu}$ at two rapidity values near mid-rapidity (y=0
and y=0.9) measured with the MRS spectrometer. The low $p_{T}$ suppression is a feature of all these 
measurements and will not be described any further, above 1.5 GeV/c, the Cronin enhancement is visible at 
mid-rapidity but its magnitude remains below 20\%. At y=0.9 there is a hint for the presence of the same
enhancement, but the extent of the pion identification range does not allow for a definite statement.
The right panel in the same figure shows a drastic change in the ratio, this time,
it grows from a value at the lowest $p_{T}$ that is consistent with $N_{part}$ scaling but remains below 
one and
reaches a value of $\sim 0.6$ at the highest $p_{T}$ bin (~3 GeV/c). This suppression in the yield of pions
measured in d+Au collisions with respect to similar yields seen in p+p collisions is well explained and 
reproduced with arguments that involve the values of the saturation scale $Q^{2}_{s}$ in the CGC formalism. 
As mentioned above, the onset of saturation, or the location of a boundary in transverse momentum space   
 where non-linear 
effects in the gluon fields are present at their full strength, depends both on the rapidity of the 
measurements and the atomic
number A of the target.  
The evolution of the gluon density, which is 
governed by equations similar to eq. \ref{eq:QuantumEvol}, will make the numerator yields change slowly with 
rapidity if the d+Au system at RHIC top energy is close to saturation. Meanwhile the p+p system at the same 
energy will have yields growing quickly with rapidity because the $p_{T}$ of the measurements is well above the
corresponding saturation scale. (The p+p system is considered dilute and is well 
described by perturbative QCD.)

\begin{figure}[!ht]
\begin{center}
   \includegraphics[height=2.in]{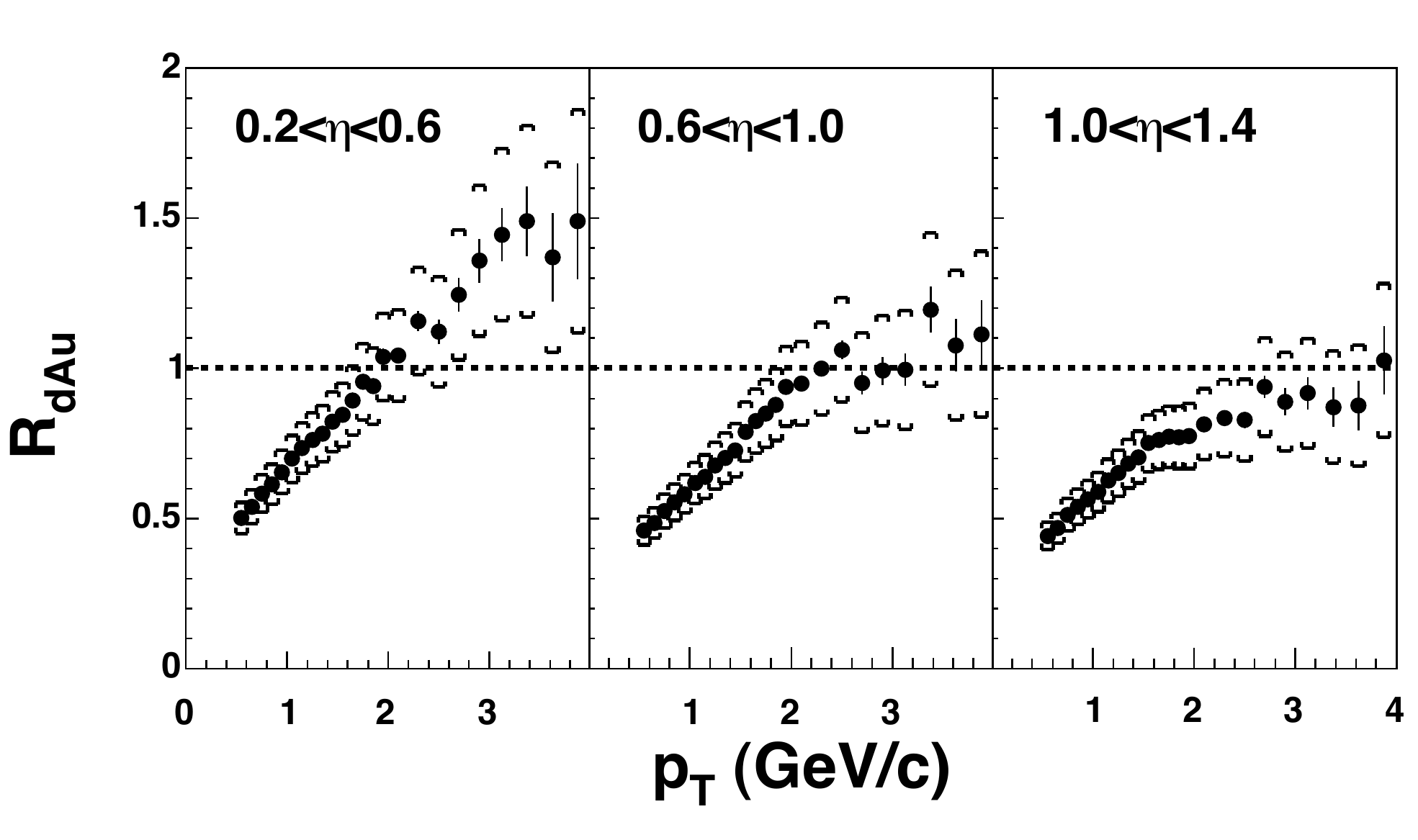}
\end{center}
  \caption{  The nuclear modification factors for charged particles measured by the PHOBOS Collaboration in d+Au
collisions and a parametrization of UA1 p+p spectra 
in three pseudo-rapidity bins close to mid-rapidity \cite{Back:2004bq}. }
  \label{Fig:phobosRdAu}
\end{figure}

The PHOBOS Collaboration has also studied the rapidity dependence of the nuclear modification factor $R_{dAu}$ with
charged particles tracks in their mid-rapidity spectrometer and their results are summarized in Fig. 
\ref{Fig:phobosRdAu} 
where they found that the suppressions starts already one unit of pseudo-rapidity away from $\eta=0$ \cite{Back:2004bq}.

The STAR Collaboration has studied the production of neutral pions  
with a lead glass array called Forward Pion Detectors (FPD) which covers a range in rapidity $3.0 < \eta < 4.2$ 
\cite{Adams:2006uz}. Their data sample includes $\pi^{0}$ with $20<E_{\pi}<55$ GeV where the energy scale
is calibrated better than 1\%.

\begin{figure}[htb]
   \includegraphics[height=2.6in]{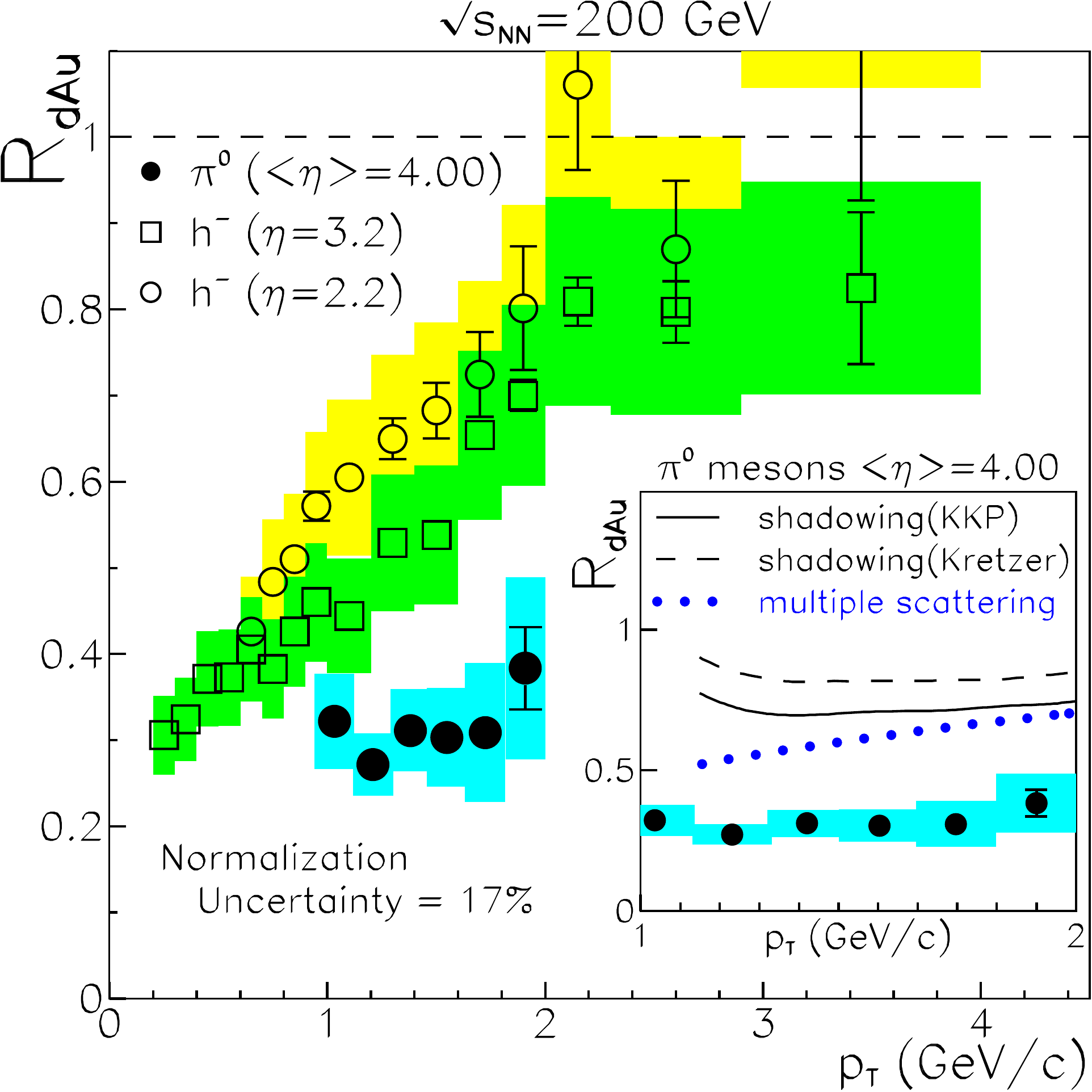}
  \caption{(color online) Nuclear modification factors $R_{dAu}$ for minimum-bias collisions. STAR
neutral pions shown with filled circles  \cite{Adams:2006uz} at an average rapidity equal to 4.00. The 
BRAHMS ratios for negative hadrons are shown with open circles and open squares at $\eta=2.2$ and $\eta=3.2$ 
respectively.  }
\label{Fig:starPi0}
\end{figure}

The minimum-bias $R_{dAu}$ extracted with $\pi^{0}$ by STAR at $\langle \eta \rangle = 4.0$ is shown in 
Fig. \ref{Fig:starPi0}
together with the published BRAHMS for negative charged hadrons at $\eta$ = 2.2 and 3.3. The suppression seen
in the BRAHMS data at high rapidity is also present in the STAR results and is even more pronounced at $\eta=4.0$.

As mentioned above, the saturation scale has an A dependence that is written more accurately as depending on
the number of participating Au nucleons \cite{Baier}:

\begin{equation}
Q^{2}_{sA}(b) = Q^{2}_{sA}(0)N_{part. Au}(b)/N_{part. Au}(0).
\label{eq:centralitySat}
\end{equation}

\begin{figure}[htb]
   \includegraphics[height=2.5in]{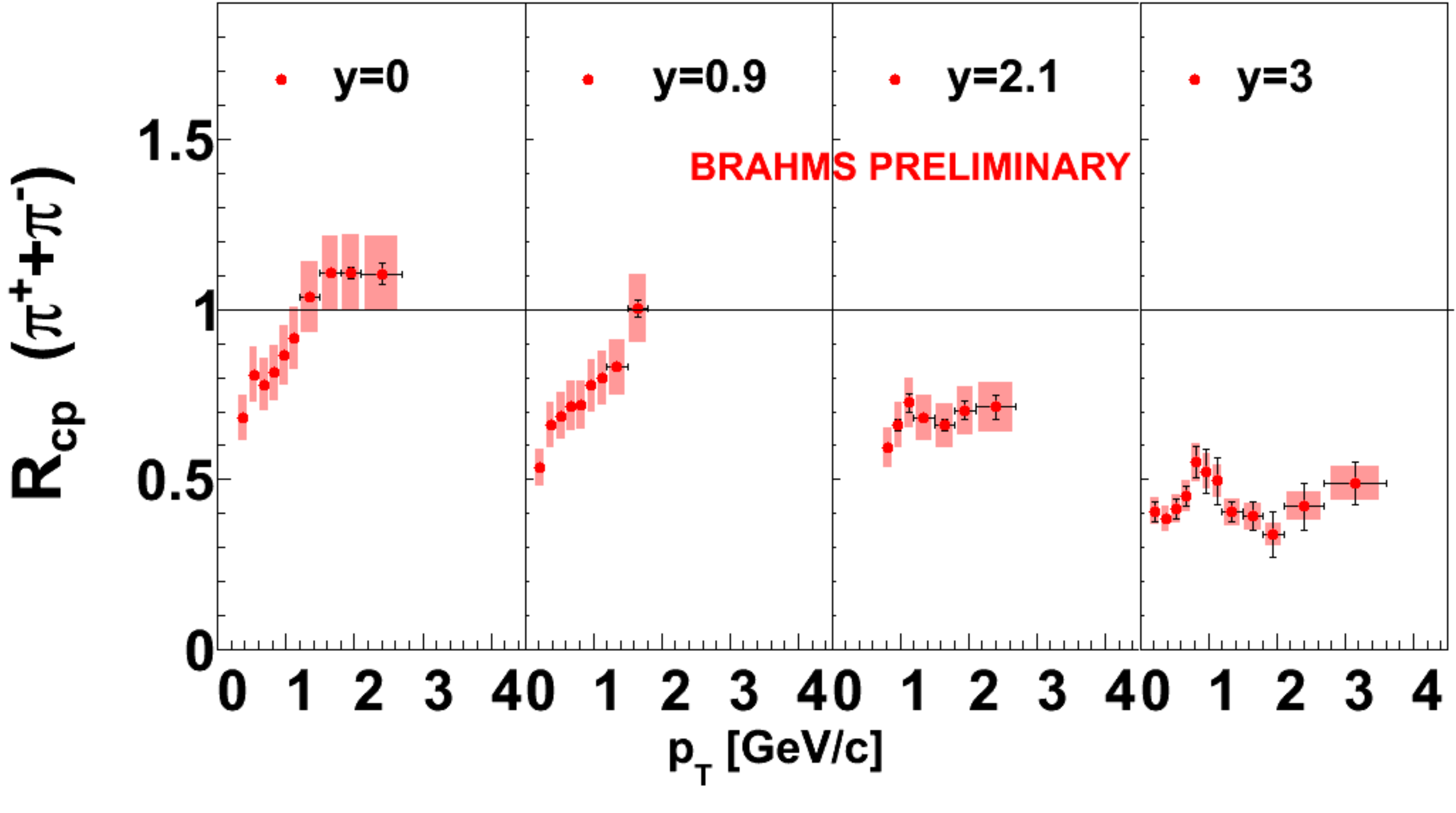}
  \caption{ The $R_{cp}$ ratio calculated with pions produced at different 
rapidities. The numerator of this ratio is populated with central events, which are normalized by the 
number of binary collisions calculated to be equal to $N_{coll} = 12.6$. The denominator is extracted 
from the 60-80\% sample of peripheral events normalized by $N_{coll} = 3.5 $.}
\label{Fig:brahmsRcpPion}  
\end{figure}

Using eq. \ref{eq:centralitySat} would imply a factor of almost 4 in the value of the saturation scale 
$Q^{2}_{sat}$ between the central and peripheral events. The effect of such increase is apparent in 
Fig. \ref{Fig:brahmsRcpPion} where central events are compared to peripheral ones with a ratio similar to
the nuclear modification factor $R_{dAu}$, but this time the p+p reference is replaced by the most 
peripheral sample of events, 60-80\%. This ratio commonly referred to as $R_{cp}$ has similar rapidity dependence 
as the $R_{dAu}$ ratio. 

\begin{figure}[htb]
   \includegraphics[height=2.5in]{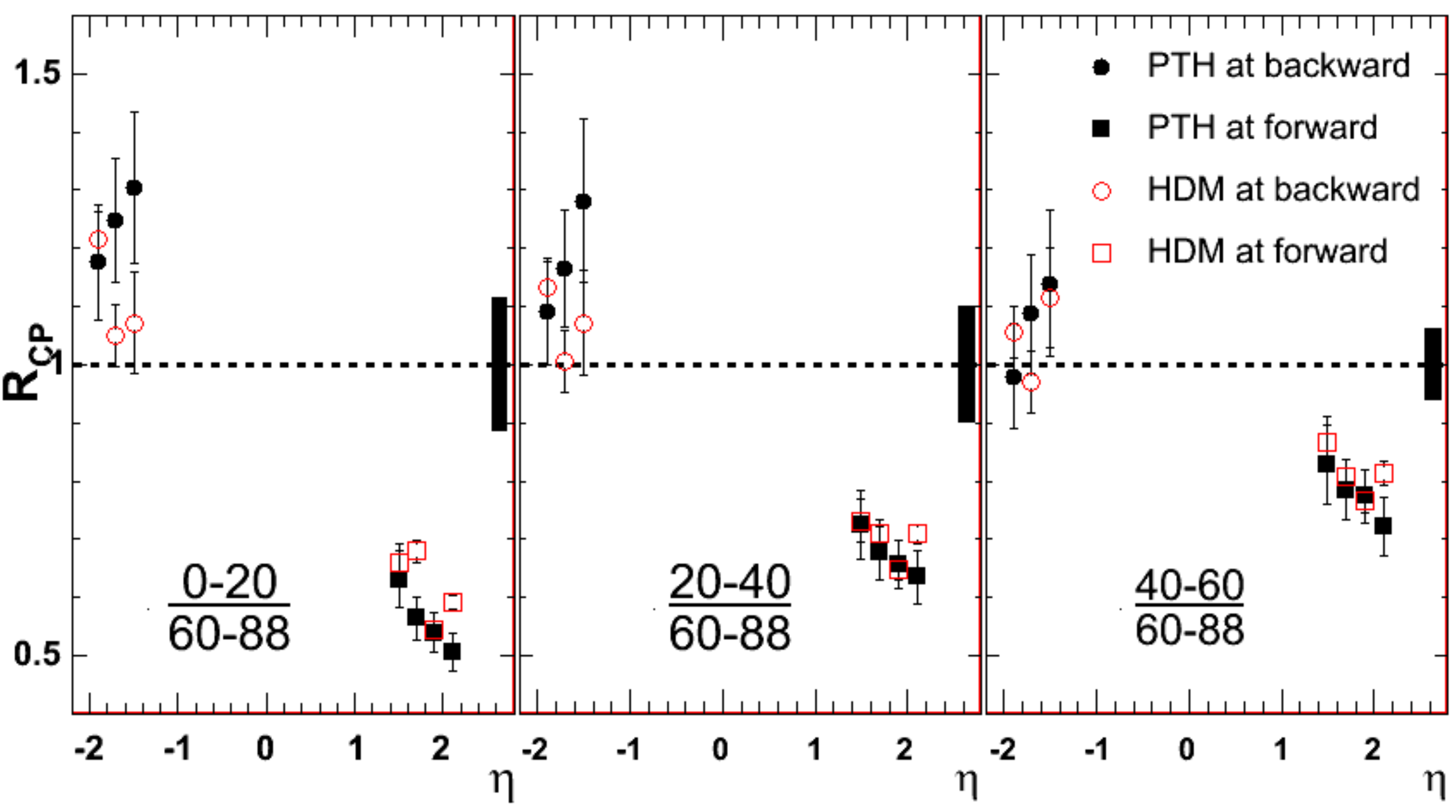}
  \caption{ Rcp of charged hadrons in the PHENIX muon arms. Show good agreement between PTH and HDM 
detection techniques. At each pseudo-rapidity they show the results integrated over 
$1.5< p_{T}<4.0$ GeV/c. }
\label{Fig:phenixPunch} 
\end{figure}

The PHENIX Collaboration has extracted the $R_{cp}$ ratio at moderately high rapidity in both projectile and target
fragmentation regions, using their Muon Arms \cite{Adler:2004eh}. The coverage in the
projectile fragmentation region (deuteron beam) is $1.4<\eta<2.0$ corresponding to $x_{2}$ in the Au target 
ranging from 0.001 to 0.03. The coverage for the Au fragmentation
is $-2.0<\eta<-1.4$ which probes $x_{2}$ from 0.04 to 0.5 where anti-shadowing  and the EMC effect  may have 
relevance. 
The measurement is based on two techniques: the first uses hadrons that make it past the muon absorbers
and reach some depth in the MuID detector, called Punch Thru Hadrons (PTH); the second technique
involves  detected muons produced by 
the decay of $\pi$ and K before
the absorber, called Hadron Decay Muon (HDM). Measurements of absolute yields are not possible, 
but the ratio of yields at different
centralities is. Figure \ref{Fig:phenixPunch} summarizes those measurements and shows a very clear difference between 
the projectile and target fragmentation regions and confirms the BRAHMS measurements at $\eta = 2$.

\section{Summary}

We have attempted to summarize the most salient measurements of the single particle inclusive spectra extracted from
d+Au collisions at \snn = 200 GeV at RHIC. The Cronin effect is present around mid-rapidity but its magnitude 
is smaller than the one seen at lower energy in fixed target measurements. The magnitude  of the Cronin enhancements 
appears to reach an upper limit as the collisions become more central. There is a clear difference between 
$R_{dAu}$ factors 
calculated with baryons and mesons.  As the  rapidity of the detected particles approaches the deuteron beam rapidity, 
the $R_{dAu}$ factors are
strongly suppressed. In the context of the CGC, that suppression is interpreted as an indication that the saturation scale at 
the relevant
values of $x_{2}$ has values in the range of the measured $p_{T}$, which in turn implies that at RHIC energies the 
wave function of the target is saturated.



\bibliographystyle{elsarticle-num}

\begin{thebibliography}{00}

 \bibitem{Cronin}J.W.Cronin {\it et al.} Phys. Rev. D{\bf 11} (1975) 3105.

\bibitem{Busza:1975fe}
  W.~Busza,
  AIP Conf.\ Proc.\  {\bf 26}, 211 (1975).
\bibitem{Busza:1989px}
  W.~Busza and R.~Ledoux,
  Ann.\ Rev.\ Nucl.\ Part.\ Sci.\  {\bf 38}, 119 (1988).


\bibitem{:2009wt}
  F.~D.~Aaron {\it et al.}  [H1 Collaboration and ZEUS Collaboration],
  JHEP {\bf 1001}, 109 (2010)
  [arXiv:0911.0884 [hep-ex]].

\bibitem{Breitweg}
J. Breitweg {\it et al.} [ZEUS Collaboration],
Eur. Phys. J C{\bf 7}, 609-630 (1999).

\bibitem{Gelis:2010nm}
  F.~Gelis, E.~Iancu, J.~Jalilian-Marian and R.~Venugopalan,
  arXiv:1002.0333 [hep-ph] and references therein.

\bibitem{Baier:2003hr}
  R.~Baier, A.~Kovner and U.~A.~Wiedemann,
  Phys.\ Rev.\  D {\bf 68}, 054009 (2003)
  [arXiv:hep-ph/0305265].

\bibitem{Kharzeev:2004yx}
  D.~Kharzeev, Y.~V.~Kovchegov and K.~Tuchin,
  Phys.\ Lett.\  B {\bf 599}, 23 (2004)
  [arXiv:hep-ph/0405045].



\bibitem{PhobosDndeta}
  B.B. Back {\it et al.} [PHOBOS Collaboration], Phys. Rev. C {\bf72}, 031901(R) (2005).
 
\bibitem{Brodsky:1977de}
  S.~J.~Brodsky, J.~F.~Gunion and J.~H.~Kuhn,
  Phys.\ Rev.\ Lett.\  {\bf 39}, 1120 (1977).

\bibitem{:2008ez}
  B.~I.~Abelev {\it et al.}  [STAR Collaboration],
  Phys.\ Rev.\  C {\bf 79}, 034909 (2009)
  [arXiv:0808.2041 [nucl-ex]].


\bibitem{:2007by}
  S.~S.~Adler {\it et al.}  [PHENIX Collaboration],
  Phys.\ Rev.\  C {\bf 77}, 014905 (2008)
  [arXiv:0708.2416 [nucl-ex]].

\bibitem{PhenixNucEff}
  S.~S.~Adler {\it et al.}  [PHENIX Collaboration],
  Phys.\ Rev.\  C {\bf 74}, 024904 (2006)
  [arXiv:nucl-ex/0603010].


\bibitem{Antresian}
  D. Antreasyan {\it et al.}, Phys. Rev. D {\bf 19}, 764 (1979).

\bibitem{Abelev:2008yz}
  B.~I.~Abelev {\it et al.}  [STAR Collaboration],
  Phys.\ Rev.\  C {\bf 78}, 044906 (2008)
  [arXiv:0801.0450 [nucl-ex]].


\bibitem{PhenixPhi}
  A. Adare {\it et al.}  [PHENIX Collaboration],
  arXiv:1004.3532.

\bibitem{PhenixOmega}
  S.~S.~Adler {\it et al.}  [PHENIX Collaboration],
  Phys.\ Rev.\  C {\bf 75}, 051902 (2007)
  [arXiv:nucl-ex/0611031].


\bibitem{Arsene:2004ux}
  I.~Arsene {\it et al.}  [BRAHMS Collaboration],
  Phys.\ Rev.\ Lett.\  {\bf 93}, 242303 (2004)
  [arXiv:nucl-ex/0403005].

\bibitem{Baier}
   R. Baier, Y. Mehtar-Tani and D. Schiff, Nucl. Phys. A {\bf764}, 515 (2006).

\bibitem{Back:2004bq}
  B.~B.~Back {\it et al.}  [PHOBOS Collaboration],
  Phys.\ Rev.\  C {\bf 70}, 061901 (2004)
  [arXiv:nucl-ex/0406017].








\bibitem{Adler:2004eh}
  S.~S.~Adler {\it et al.}  [PHENIX Collaboration],
  Phys.\ Rev.\ Lett.\  {\bf 94}, 082302 (2005)
  [arXiv:nucl-ex/0411054].
\bibitem{Adams:2006uz}
  J.~Adams {\it et al.}  [STAR Collaboration],
  Phys.\ Rev.\ Lett.\  {\bf 97}, 152302 (2006)
  [arXiv:nucl-ex/0602011].

 \end{thebibliography}




\end{document}